\documentclass{jpsj-suppl}
\usepackage{txfonts} 

\title{Challenges in Cosmology 
from the Big Bang to Dark Energy, Dark Matter and Galaxy Formation}

\author{Joseph \textsc{Silk}$^{1,2,3}$ }
\inst{$^{1}$Institut d'Astrophysique de Paris, Sorbonne Universit\'e, UPMC Univ. Paris 6 \& CNRS, UMR 7095,  98 bis Boulevard Arago, F-75014 Paris, France\ \\
$^{2}$Department of Physics and Astronomy, 3701 San Martin Drive, The Johns Hopkins University, Baltimore MD 21218, USA\ \\
$^{3}$BIPAC, 1 Keble Road, University of Oxford, Oxford OX1 3RH, UK}

\email{jpsj{\_}edit@ipap.jp}
\email{silk@iap.fr}

\recdate{\today }

\abst{I review the current status of Big Bang Cosmology, with emphasis on current  issues in dark matter, dark energy,  and galaxy formation. These topics motivate many of the  current goals of experimental cosmology which range from targeting the nature of dark energy and dark matter to probing the epoch of the first stars and galaxies.
}
\kword{cosmology, dark energy, dark matter, galaxy formation}

\begin{document}
\maketitle

\section{Introduction}
Cosmology has entered a new era in the twenty-first  century. It has become a precision science. Nevertheless, big questions remain. I will attempt to address some key issues in this talk, focussing on these topics:
\begin{itemize}
\item The Big Bang
\item Dark Energy
\item Dark Matter
\item Galaxy Formation\
\end{itemize}

Modern cosmology began with Alexander Friedmann in 1922 and especially with Georges Lemaitre, who in 1928 sketched  the current suite of  cosmological models and realized that there was indeed tantalizing evidence for expansion of space from Vesto Slipher's measurements of galaxy redshifts. In 1929, Edwin Hubble, indeed unaware of the earlier work, announced  his discovery of the law that relates redshift linearly to increasing distances of the galaxies, and the expanding universe was born. Lemaitre himself was aware of the problems of formally following his equations back to the singular state $t=0$, and advocated the "primeval atom" as a more physical starting point. The Lemaitre-Eddington solution, motivated  to avoid the initial singularity, starts from a static configuration but  was later found to be unstable. This set the scene for  inflation along with other possibly less-inspiring alternatives. After nearly 90 years of progress in observational cosmology, all that has really changed on the theory side is the addition of inflationary cosmology to our cosmic toolbox,  allowing the extension of physical  cosmology back to the Planck time, some $10^{-43}$ seconds into  the Big Bang.

\section{History  of physical cosmology1933-2000}

\subsection{ The standard model}
Little new happened in our understanding of cosmology for  the first two decades following Hubble's announcement of the recession of the galaxies, and its rapid incorporation into evidence for the expansion of space.
It was to take the entry into cosmology of nuclear physicists  to allow the next giant step forward. This was a consequence of the realization that the Big Bang was an ideal laboratory to study nuclear physics, and in particular the awareness  that the high density and temperature achievable in the first minutes might facilitate the nucleosynthesis of the chemical elements. 

George Gamow dreamt that all of the chemical elements could be synthesized  primordially, but he eventually realized that  the lack of a stable element of atomic mass 8 made it impossible to go beyond lithium (mass 7). With Alpher and Herman, he did however succeed in generating helium, the second most abundant element in the universe, in the Big Bang. This  was a major accomplishment, as it was realized in the 1950s that ordinary stars could not generate enough helium as observed in nearby stars \cite{tayler}. Nor could stars account for what we now know to be its uniformity in abundance throughout the observable universe. It was soon shown in a classic paper by Burbidge, Burbidge, Fowler and Hoyle, that the ejecta of massive stars exploding as supernovae, as well as red giant  mass loss, could account for the bulk of the heavy elements \cite{burbidge}.

Modern calculations of primordial nucleosynthesis also account for the abundance of deuterium, an isotope of hydrogen which is only destroyed in stars. Lithium is 
also produced, but the amount predicted is about 3 times more than observed in the oldest stars \cite{coc}. The cosmological origin of lithium is one of the very few  serious questions that are as yet unresolved in cosmology. Another important consequence of helium and deuterium synthesis is that 
the bulk of the universe is nonbaryonic. If the bulk of the matter in the universe were baryonic, the nuclear reactions in the first few minutes would have been so efficient as to overproduce helium
by a large amount.  However a  quantitative baryon budget was to require Hubble Space Telescope observations of intergalactic and circumgalactic gas  in order to measure the total density of matter with high accuracy, e.g. \cite{burles}.

On the observational side, the cosmic microwave background radiation was to provide  a basis for undertaking the first precise measurements of the different components of the universe. In fact, precision cosmology emerged after 2000 as a consequence of several generations of cosmic microwave background experiments, following its detection by Penzias and Wilson in 1964 \cite{penzias}, and the measurement of its blackbody spectrum by COBE in 1990 \cite{mather}.

The major theoretical development in cosmology, following the glory days of Lemaitre, de Sitter, Einstein and Eddington in the 1930s, was the invention of inflationary cosmology, attributed to Starobinsky, Guth and Linde  \cite{starobinsky, guth, linde}   although others played important roles, e.g.  \cite {sato,  albrecht}.
Inflation accounted for   the flat geometry of the universe, its size and for the presence of infinitesimal density fluctuations.  The latter were initially found to be too large in the early inflationary models, but  a consistent theory was first developed by Mukhanov \cite{mukhanov}. The inflationary density fluctuations were predicted to be adiabatic, gaussian, and nearly scale-invariant. All of these properties were later confirmed to a high degree of approximation by the WMAP and Planck satellites.

\subsection{Dark matter}
The case for dark matter in galaxies was effectively made quite early, with optical studies of M31 by Rubin \& Ford in 1970  \cite{rubin}, and especially with the first deep 21cm observations of M31 and other nearby galaxies by Roberts and others in 1972 \cite{roberts}, although the argument for the prevalence of dark matter in clusters of galaxies was made decades earlier by 
 Zwicky in 1933 \cite{zwicky}. 

Far more matter is needed than is allowed in the form of baryons, by a factor of about 6. A higher baryon density would  have generated  excessive helium and too little deuterium. There was another  mini-revolution in cosmology following the discovery of inflation in 1981, when it also became clear that the early universe was a fertile hunting ground for novel dark matter-related  ideas in high energy particle physics. Within two or three  years, a number of particle physics candidates emerged for nonbaryonic dark matter. Now there are hundreds of candidates, but there are still  no detections.

Two preferred candidates attracted most attention. One was the lightest neutralino, a left-over particle from the epoch when supersymmetry (SUSY) prevailed, proposed by  Pagels \& Primack in 1982 \cite{pagels}.  Neutralinos are an attractive candidate because if the natural scale  for the breaking of SUSY is that of the W and Z bosons, around 100 GeV, one naturally obtains the observed relic density of dark matter.  

Many experiments were launched to search for neutralinos in the form of weakly interacting massive particles (WIMPs), inspired by the possibilities of direct detection via elastic or even inelastic scattering \cite{goodman} and of indirect detection, via the annihilation products that include high energy antiprotons, positrons and  gamma rays  \cite{srednicki} and neutrinos \cite{sos}.
 Sadly, accelerator experiments have frustrated this dream by failing to find evidence for SUSY below a TeV.

 A second candidate came to be dubbed the invisible axion. Axions were invented to solve an outstanding issue in QCD, the strong CP problem,   by Wilczek \& Weinberg in 1978 \cite{weinberg, wilczek }. Neutralinos and axions  are intrinsically  cold dark matter (CDM), that is, dark matter which can respond to the pull of gravity on all mass scales as advocated by  Blumenthal et al. in 1984 \cite{blumenthal}.  
 
 Cold dark matter was  shown to account for large-scale structure by Davis et al. in 1985 \cite{davis}, in a way that hot dark matter (HDM), could not accomplish. This was epitomized by massive neutrinos and effectively nonbaryonic matter created late in the universe  with a significant velocity dispersion,  hence said to initially be hot, could not accomplish. Warm dark matter, perhaps in the form of sterile neutrinos, remains a possible option that does not affect large-scale structure, unlike HDM, but improves predictions at dwarf galaxy scales via enhanced free-streaming relative to CDM  \cite{bond83}.
 
\subsection{Acceleration}

The case  for acceleration of the universe from Type 1a supernovae was  first presented in 1998  by Perlmutter, Riess, Schmidt and colleagues \cite{perlmutter, riessa}. These results are interpreted as strong evidence for the  cosmological constant, itself considered to be evidence for the presence of a dominant component of dark energy.  The flatness of the universe, measured by CMB experiments such as BOOMERANG, was a notable factor in establishing the need for dark energy as well as for dark matter \cite{lange}.
It was not until 2000 that scientists  decisively showed that the universe contained a near-critical density of mass-energy by measuring the curvature of space via the CMB. It became clear  that the dominant energy-density component  was not predominantly dark matter. Had it been, galaxy peculiar velocities would have  been  excessive because of the clustering effects of  dark matter. Dark energy dominates as a scalar field, and remains  nearly uniform,  acting like antigravity on the very largest scales.

\subsection{Galaxy formation}
After  robustly developing the framework for the Big Bang, the next challenge was to understand how galaxies, and more generally large-scale structure, formed. The seed fluctuations needed to form structure by gravitational instability in the expanding universe were first predicted  in 1967 as temperature anisotropies in the cosmic microwave background  \cite{silk} that damped out towards smaller scales  \cite{silka, wilson}. In fact 
the first rigorous calculations of density irregularities in a coupled matter-radiation fluid  \cite{peebles} identified what were later recognised as the baryon acoustic oscillations \cite{eisenstein}.
%
The early calculations were later improved  by incorporating the growth-boosting effects of dark matter, leading to smaller fluctuations 
\cite {vittorio, bond}.
It  was to take more than another  decade for the detection of these elusive temperature fluctuations. The first indications   came from
the COBE satellite in 1992, but the measured temperature anisotropies were only on 7 degree angular scales. These are super-horizon at last scattering and demonstrate the imprint of quantum fluctuations arising at inflation. 

However much finer angular scales needed to be studied in order to seek  the direct  traces of the seeds of the largest  self-gravitating structures such as clusters of galaxies. These  seeds produce cosmic microwave background fluctuations   on  angular scales of a few arc-minutes. These fluctuations were only accurately measured a  decade later, as the unique features of the predicted acoustic sound wave damping peaks, detected by  the WMAP satellite. Indeed this telescope detected the first three peaks,
and after another decade, the SPT and ACT telescopes  measured a total of seven  damping wiggles in 2011 \cite{keisler, dunkley}. There was no longer any doubt that we were seeing the seeds of structure formation imprinted on the microwave sky.

Observational surveys soon became the preferred tool for low redshift  cosmology, that is cosmology over look-back times 
of up to ten billion years.  Million galaxy surveys, notably 2DF (1997) and SDSS (2000), set out the observational basis for a cold dark matter-dominated universe. The next step towards connecting primordial fluctuations with galaxies required  detection of the acoustic imprint of the primordial  seed fluctuations on the matter distribution  \cite{sz}. 
Improved, more accurate, spectrocopically calibrated  surveys, were needed for  detection, which eventually came with the next generation of galaxy redshift surveys, that would be aimed  towards a redshift of order unity..

Growth from  seeds to structure  formation involved the operation of gravitational instability. This operated once the radiation decoupled from the matter as the universe recombined 380,000 years after the Big Bang. Structure formation was greatly aided by the self-gravity of the dark matter. The case for cold dark matter driving galaxy formation with the aid of baryonic dissipation was made by White \& Rees in 1978 \cite{white}.

It also rapidly became apparent that  complex baryonic physics was needed in addition to the effects of gravity  to result in structures that resembled the observed distribution of galaxies.  For one thing, the intergalactic gas had to condense and cool in order to make the dense cold clouds that can fragment and form stars.  It soon became evident that the physics of supernova feedback was required to avoid excessive formation of large numbers of dwarf galaxies \cite{dekel}. A similar problem for massive galaxies was avoided by the presence of supermassive black holes, releasing prolific amounts of energy in a quasar phase contemporaneously with the formation of the oldest stars, that left its imprint on the relation between black hole mass  and spheroid velocity dispersion \cite{reess}.

\section{Modern cosmology}
Theory has progressed enormously  in the past decade, especially in the area of simulations of large volumes of the universe. While massive large-volume, large-scale structure simulations provide an impressive match to survey data, real issues remain however with regard to  small-scale cosmology, namely galaxy formation.
Of course to explain the late universe requires a lot of dirty physics. 
It is a bit like predicting the weather. But the early universe  is much simpler. Let me first summarize the current status of precision cosmology from the CMB.

\subsection{Cosmic microwave background radiation}

All-sky mapping of the CMB has proven to be crucial in evaluating the parameters of cosmology with high precision. This has been the important contribution of the WMAP and especially the Planck experiments  \cite{planckXIII}. Seven 
numbers  characterize the  CMB fluctuations (if one does not assume a spatially  flat universe).
Apart from the densities of dark energy $\Omega_\Lambda$, dark matter $\Omega_m$ and baryons $\Omega_b$, one has the scalar spectral index $n_s $ and  normalization $\sigma_8$ of the density fluctuations, the (revised) optical depth $\tau$ \cite{PlanckXLVI} and the Hubble constant $H_0$ (Table I). The inferred age of the universe is 
13.799 $\pm$ 0.021 Gyr. 

There is general  consistency with  other probes, notably baryon acoustic oscillations (BAO) and  weak gravitational  lensing, with some slight tension remaining in the latter case.
A slight tension remains at the $2\sigma$ level with the slightly higher Planck 2015 determination of the fluctuation amplitude parameter $\sigma-8$ \cite{hildebrandt2017}.
There is somewhat more tension with the Hubble  constant measurements, found locally to be slightly higher than the Planck 2015-preferred value: $73.24\pm 1.74$ km/sec/Mpc   \cite{riess}.

These differences are possibly due to  systematics between  the very different distance scales probed, although more profound explanations cannot be excluded. It is already quite remarkable how well the very early universe as mapped by Planck matches the local universe, as charted via weak lensing, BAOs and supernovae, in the context of the standard model of cosmology, $\Lambda$CDM.

 B-mode polarization is predicted to be a unique probe of the gravitational wave background imprinted by inflation \cite{kamionkowskiP, seljak}.  Many experiments are underway to address this goal. Most notably, spectral distortions in the blackbody spectrum arising  from the standard model of structure formation due to dissipative processes are predicted to be
in the form of a negative chemical potential distortion  at a level of a part in $10^8$ \cite {hu94}. Detection of $\mu$-distortions at this level would probe fluctuation damping  and primordial non-gaussianity on  dwarf galaxy scales \cite{pajer2012}.
However such values of $\mu$  are four orders of magnitude below the COBE FIRAS limits, and only detectable with a dedicated future space experiment \cite{sunyaev}.

\begin{table}[tbh]
\caption{Cosmological parameters: 2016\cite{planckXIII,PlanckXLVI}}
\label{t1}
\begin{tabular}{ll}
\hline
$\Omega_b$ & $0.0486\pm  \ 0.0005 $\\
$\Omega_m$ & $0.3089\pm  \ 0.0062 $\\
$\Omega_\Lambda$ & $0.6911\pm  \ 0.0062 $\\
$\sigma_8$ &$ 0.8159 \pm 0.0086$\\
$n_s $ & $0.9667\pm 0.004$\\
$H_0$ & $67.74\pm 0.46$\\
$\tau$ & $0.056\pm 0.009$\\
\hline
\end{tabular}
\end{table}

\subsection{
Baryon acoustic oscillations and weak gravitational lensing}

The large-scale structure of  the galaxy distribution is a fertile hunting ground for evaluating both the geometry of the universe and its rate of expansion \cite{percival}. The baryonic acoustic oscillation  (BAO) scale  $\sim 150 $ Mpc is a standard ruler for cosmology.  It is the precise analogue of the photon oscillations at $z=1080$ on the  last scattering surface  but is observable  in the large-scale distribution of the galaxies  at the present epoch and out to $z\sim 1$ \cite{beutler16}.   Its remarkable power is that measurement requires no knowledge of the type of tracer. Comparison of the two length scales constrains the angular diameter distance  and Hubble parameter at different redshifts. Hence deviations in dark energy due to possible variations in  the cosmological constant assumption can be inferred.  

The angular diameter distance measurements are especially sensitive to the geometry of the universe.
Weak gravitational lensing, galaxy clustering and redshift space distortions all measure the dark matter content of the universe to high precision, and at different redshifts constrain  the growth rate of density fluctuations and its dependence on dark energy. Supernova measurements constrain the expansion rate of the universe. Degeneracies persist between cosmological parameters, leaving scope for dynamical dark energy or modified gravity, to be probed by missions such as EUCLID \cite{euclid}.

\subsection{Dark energy}
No deviations with look-back time are detected  from  the cosmological constant $\Lambda$  in  the standard model to $< 5\%$.
The small value measured for $\Lambda$ represents what has been called the greatest problem in physics.
There are two key questions: 
why is  $\Lambda$ so small,  by $\sim $120  powers of 10 as compared to the vacuum density at the scale of grand unification? And why is it just becoming dominant today, 
at  $\sim$$ 10^{17}$  second rather than  at grand unification some $10^{-36}$  second after the Big Bang?
Is the explanation due to particle physics or due to astrophysics? Or do we need to wait for a fundamental theory  of quantum gravity to emerge?  

There is a particle physics "explanation", or rather, the indication of a possible route to be further explored. Consider a topological classification of all Calabi-Yau manifolds. This  idea seems promising, in reducing  more than $10^{500}$ manifolds  predicted by string theory to a much smaller number,  those with small Hodge  number, but  needs further exploitation in terms of phenomenological string theory to be a useful guide \cite{candelas16}.  Nevertheless, this approach may provide a hint as to an eventual possible selection principle  for a universe congenial to both the standard model of particle physics and a low value of the cosmological constant.

And there is an astrophysics "explanation",  or again, more realistically,  a  hint of a promising direction that merits an in-depth study. Consider a universe with a few very  large voids. Putting us near the center of a huge Gpc-scale  void could explain our apparent acceleration, but at the price of producing excessive peculiar velocities and excessive kSZ signals \cite{garcia12}. A more controversial approach is to postulate a universe with many large voids whose cumulative effect  via back reaction may arguably  contribute towards global acceleration, albeit probably not  at a sufficient level to account for dark energy
\cite{buchert16}.

\subsection{What is dark matter?}
Dark matter is not baryonic. This is a canonical  result, emanating from nucleosynthesis considerations in the first minutes of the Big Bang. Confirmation comes from 
the observed abundances of helium and deuterium, as well as the inferred effective number of neutrino species, $N_{eff}= 3.15 \pm 0.23,$ versus    the predicted value of 3.046.
This  differs from a pure integer because neutrino decoupling precedes electron-positron pair annihilation but is close enough so  that neutrino flavour oscillations generate neutrino spectral distortions that  slightly enhance the number density of relic neutrinos \cite{mangano}. 

Primordial
nucleosynthesis is highly successful in accounting for light element abundances, albeit with a question mark 
over Li overproduction. The matter fraction (including dark matter)  is 
$\Omega_m = 0.3089 \pm  0.0062,$  whereas the baryon fraction, with $h=H_0/100\rm km^{-1}s^{-1}Mpc^{-1},$  is  $\Omega_b h^2 = 0.02230 \pm  0.00014.$  
Hence 
84\% of the dark matter 
is not made of baryons. 

Nor is it made of standard model neutrinos. The sum of neutrino masses required to account for dark matter is 
$\Omega_\nu h^2 = \sum m_\nu/93\rm{eV},$  whereas from the Planck data $ \sum m_\nu < 0.23$eV. Including limits from the Lyman-$\alpha$ forest of intergalactic hydrogen clouds extends the scales probed to smaller comoving mass scales, albeit at the risk of introducing systematics associated with hydrodynamic modeling of the intergalactic medium, to set a  limit of $ \sum m_\nu < 0.12$eV \cite{palanque}.
The measured contribution from neutrino oscillations is about 58meV, so that $\Omega_\nu h^2 > 0.0006.$ The maximum value is $\Omega_\nu h^2 < 0.0025.$

The mystery of dark matter is that despite intensive searches, both indirect \cite{gaskins16} and direct \cite{marrodan16}, its nature has not been identified.
There is little question as to its observational dominance on scales from tens of kpc up to horizon scales, assuming Einstein gravity.
It  most likely consists of  massive weakly interacting particles. The  most attractive ansatz for the WIMP particle has long been provided by supersymmetry, which motivates the so-called WIMP miracle. If the lightest stable SUSY particle was once in thermal equilibrium, 
$<\sigma v>_x \sim 3.10^{-26}\rm {cm^3 s^{-1} } \sim 0.23 / \Omega_x,$ with the consequence that  thermal WIMPs  generically account for the dark matter if the cross-section is typical of weak interactions.  There is a wide range of SUSY models that give scatter in the cross-section at fixed WIMP mass by several orders of magnitude while still giving the correct dark matter density. The 100 or more free parameters in SUSY have led theorists to explore minimal models \cite{cirelli2011}, which are highly predictive and invaluable for guiding indirect detection experiments,  but at the price of  restrictions on the full range of dark matter candidate masses and annihilation channels.

Fermi satellite observations of an excess, relative to standard templates, diffuse gamma ray flux in the galactic center region have motivated a mini-industry of dark matter interpretations, usually involving WIMP self-annihilations. The preferred WIMP mass is around 40-50 GeV, although this depends on the adopted annihilation channels  \cite{daylan}. The dark matter interpretation requires a thermal cross-section, as favored by WIMP freeze-out in the very early universe. An alternative interpretation involving gamma rays from an old population of  millisecond pulsars has received much attention, and is even favored by recent studies of evidence via fluctuation analyses for  a faint discrete source population \cite{lee}. 

High energy positron and antiproton measurements by, most recently, the AMS-02 space experiment have also been claimed to provide evidence for dark matter particles of mass near a TeV.  There is a new component of positrons beyond expectations from secondary cosmic ray production.  However, the interpretation of the rising  positron/electron ratio with increasing energy as a dark matter signal requires an implausibly high annihilation signal, and an astrophysical interpretation in terms of  $e^+e^-$ pairs produced by nearby pulsar winds seems more likely, cf.  \cite{dimauro}. The observed antiproton flux is consistent with secondary production \cite{cirelli2015}.

As the WIMP mass is  increased, the scaling 
$<\sigma v>_x \sim\alpha_W^2/m_x^2 $ indicates that there is a maximum WIMP mass to avoid overclosing the universe, the so-called unitarity limit, of 
around 50 or 100 TeV \cite{kamionkowski}. The possibility of exploring such a large mass range, being near the limit of most acceptable SUSY models, is a prime reason for encouraging the construction of a future 100 TeV collider \cite{arkani16}. 

Candidates for cold dark matter include neutralinos, axions, and primordial black holes. 
Warm dark matter is another viable option, characterized most recently by 
7 keV sterile neutrinos as motivated by controversial claims of a 3.5 keV  x-ray line in the Perseus cluster and elsewhere. This particular candidate, just the most recent of a long series, seems to have been largely abandoned as of 2016 \cite{shah, profumo}.

The large-scale structure of the universe is well reproduced by cold or warm dark matter. The progress  in surveys and in numerical simulations has been noteworthy. The first galaxy redshift survey  (the CfA survey) in 1981 was of around 1000 galaxies, and was modeled with simulations of   32K particles  \cite{davis85}. Surveys and simulations advanced rapidly.
After two decades, both large data  surveys  (the 2dF and Sloan digital galaxy redshift  surveys of $\sim 10^6$ galaxies) and  simulations  with $\sim 10^8$ N-body masses  were accomplished.
Current simulations can cope with a  trillion particles  (as of 2016).  

However addition of baryonic physics is crucial if we are  to go beyond using mass points as tracers of large-scale structure. This remains a controversial subject in view of the  differing prescriptions for the necessary subgrid physics.

\section{Advances in Galaxy Formation}

Some 380000 years after the Big Bang, temperature fluctuations imprinted on the last scattering surface provide evidence for the  spectrum of primordial adiabatic fluctuations that seeded galaxy formation about a billion years later. Gravitational instability of weakly interacting  cold (or warm) dark matter enables the growth of structure in the form of galactic halos that culminates  in the formation of the first  dwarf galaxies. Gas accretion triggers star formation, and the subhalos merge together as gravitational instability relentlessly continues until the full range of galaxy masses has developed.

But this grand scenario hides the ad hoc  subgrid physics. The 
evolution of massive gas clouds is   controlled by the 
competition between self-gravity and atomic  cooling. The building blocks of galaxies are  clouds of $\sim 10^6$ solar masses, since these form at a redshift of order 20 and are the smallest  
self-gravitating clouds that are warm enough to excite trace amounts of $H_2$ cooling and allow the first stars to form \cite{palla83}.

The largest galaxies weigh in at about $\sim 10^{13}\rm  M_\odot$  in the cold dark matter theory, and form as the dwarf galaxies  merge together hierarchically in a bottom-up fashion, as dictated by the approximate scale-invariance (measured via the CMB)  in Fourier space of the primordial spectrum  of density fluctuations. One success is that 
simple scaling physics can explain the characteristic mass of a galaxy ~$\sim 10^{12}\rm  M_\odot , $  as this is the largest halo that can effectively dissipate gas energy and form stars
\cite{rees, silkb}.

Numerical simulations are used to explore the detailed process of galaxy formation, from that of the first stars to massive galaxies.
However we cannot resolve star formation or AGN feedback in cosmological zoom-in simulations. The best one can do is to resolve parsec scales, in state-of-the-art trillion particle simulations. However feedback physics is set on microparsec scales. The usual compromise is to model sub-grid physics  by local observational phenomenology. The current generation of simulations includes both supernova and AGN/quasar  feedback, using jets and/or mechanical outflows. Many aspects of the observed galaxy population can be explained. But some cannot. No model yet accounts for all aspects of galaxy formation.

For example, an exploration of various supernova feedback recipes found that no current prescription was able to reconcile the observed star formation efficiency per unit of cold gas in star-forming Milky Way-type galaxies with the mass-loading observed for galactic outflows in a wide range of models  \cite{rosdahl}. Improvements might include. among others,  more sophisticated turbulence prescriptions \cite{semenov},  AGN-loaded superbubbles \cite{keller} or cosmic ray pressure-induced cooling delays \cite{salem}.

More globally, dwarf galaxy issues, including their predicted numbers, the question of cores versus cusps, and the "too big to fail" problem, have been largely resolved by improved baryonic physics and high resolution simulations \cite{fattahi}. However even here, not all studies agree on whether supernova feedback is sufficiently effective in a multi-phase interstellar medium in generating the  low stellar content of the ultrafaint dwarfs\cite{bland15} or in producing the recently discovered population of ultradiffuse galaxies  in both cluster and group environments \cite{vandokkum2015, merritt2016}. Nor is it obvious how to  reconcile the  history of inefficient star formation and baryonic mass loss with the efficient chemical seeding observed  of r-process elements \cite{ji16}
observed in at least one ultrafaint dwarf galaxy.   Moreover, the AGN/high star formation rate connection at high redshift continues to be an enigma, if galaxy mergers are indeed not a general panacea for this correlation \cite{fensch}. More complex feedback physics,  both negative and positive, may be needed than has hitherto been incorporated into the simulations.

Some authors are sufficiently desperate that they favor warm \cite{bose16},
fuzzy \cite{marsh14, hui16} or even self-interacting \cite{schneider16} dark matter as a general panacea for dwarfs, although these particle physics solutions are highly debated.  For example, removing small-scale power has negative consequences at high redshift, especially for early galaxy formation \cite{maccio12, maccio13} and reionization of the universe \cite{bozek15}.

As observational  facilities are built to probe ever higher redshifts, doubts persist on the physics of galaxy formation.
The state-of-the-art numerical simulations of galaxy formation are beautiful,  but are they reliable, and more specifically, are they predictive? The simulations  are  becoming increasingly complex, in terms of input physics. One can easily  imagine that they will soon be as difficult to analyze as the observational data, in terms of extracting any fundamental understanding of how galaxies actually formed.

\section{The Future for Cosmology}

The CMB temperature fluctuations have ushered in the modern age of precision cosmology. The low optical depth measured by Planck means that understanding reionization is much less of a challenge for theoretical models. The
current goal that dominates most proposed experiments  is to probe inflation via its 
gravitational wave imprint. Neutrino mass measurements are an important corollary, but will never replace direct experiments. Polarization searches for the B-mode imprint of inflation are confused by the foreground B signal, most notably from galactic dust.  The immediate goal is to increase the sensitivity. The Planck satellite was good for microKelvin sensitivity. 

To achieve B-mode polarization detection, one needs to reach the nanoKelvin level.
Several  ground-based experiments are underway using thousands or eventually tens of thousands of bolometers, or being planned  in the CMB Stage IV effort \cite{abazajian2016}
with hundreds of thousands of pixels, in contrast to the 32 bolometers on the Planck HFI instrument. Space has other advantages, with regard to overcoming systematic errors and foregrounds, especially at large angular scales, and successors to Planck are also being designed. 

One can expect the best  CMB measurements to probe $N\sim \ell^2\sim  10^6 $ independent Fourier modes on the sky. These  independent  samples to $\ell\sim 1000$ allow one to  achieve $N^{-1/2} \sim $0.1\% precision on large scales, corresponding to the precursors of galaxy clusters, before damping of the primordial fluctuations  sets in and eliminates any smaller scales from detectability on the last scattering surface.

Galaxy surveys  allow one to approach smaller scales and increase the independent sample number, and hence precision. Moreover, one gains by working in 3-D, with redshift information. Future surveys should allow
$N~\sim 10^8$  independent samples. This could provide an order of magnitude increase in precision over CMB measurements for cosmology. However galaxies are messy and biased. Larger numbers are surely needed  per independent sample. Hence their advantage over the relatively clean CMB is limited by bias and systematics.

There is only one option to truly enhance the accuracy of future cosmology experiments. 21 cm probes of the dark ages provide the ultimate precision by sampling 
Jeans mass HI gas clouds at  $z\sim 30$ that are still in the linear regime \cite{loeb05}. These clouds are colder than the CMB, as long as we observe them before galaxies or quasars, or even the first stars,  have formed, that is at very high redshift.

Using 21 cm absorption against  the CMB  allows one to approach a very large number of independent modes, perhaps $N \sim 10^{10}.$    This would allow 100 times better precision than attainable with the CMB. One can now imagine trying to measure a guaranteed prediction of inflation, namely primordial nongaussianity. This is generically expected in inflation, and is measured by  nonvanishing higher order correlations.
With sufficiently high precision, one could attempt to measure the fluctuation bispectrum.  Primordial non-gaussianity  on all scales is a more generic but more challenging and elusive prediction of inflation. This would be an important probe of inflation, and  indeed is actually guaranteed, albeit at only a low level (the usual measure of second-order nongaussianity $f_{NL}\sim 1-n_s\sim 0.02$) in single-field slow roll inflation  \cite{maldacena}.

There are many inflationary models, involving multiple inflation fields  or inflationary features  in the primordial power spectrum of density fluctuations \cite{renaux}, where effects of scale-dependent nongaussianity could be substantial, especially if enhanced small-scale power is present that would not be constrained by the Planck  limits $f_{NL}\sim \mathcal{O}(\pm 10) $  \cite{planckNG} on larger scales.

To attain such a large number of modes at high redshifts, ideally in  the redshift range 30-60, 
 one would need to use 21cm measurements at  low radio frequencies, below 100 MHz, combined with high angular resolution. These requirements necessitate a large  array of  dipoles  in a telescope site in an area with low radio interference. For example, a dipole array would need to be of size $\ell\lambda/2\pi\sim 300$ km in order to resolve Fourier modes $\ell\sim 10^5$ or a few arc-seconds 
corresponding to a cloud size of order 1 Mpc and a corresponding  bandwidth of 0.1 MHz \cite{munoz}. Perhaps only the far side of the Moon would provide a suitable site. This would be  a project to consider for the next century.

\section{Summary}

There has been remarkable progress in cosmology over the past 2 decades, and it is
now a precision science. However  there are still  important questions  to be addressed. Here is a personal subset.
\subsection{Theory}
I begin with theoretical questions
\begin{itemize}
\item 
Dark matter is here to stay but what is it?
Should we  change our theory if we fail to detect dark matter within a decade, or two?

\item 
Does the low value of dark energy requires  fine-tuning? And does this justify appeal to the multiverse with so many free parameters and little  prospect of verifiability?

\item Does dark energy vary with epoch? This might be the simplest way to avoid appeal to anthropic arguments  to understand its low  value at the present epoch.

\item
Is galaxy formation more of the same at early gas-rich epochs?
Or is something radically new needed? Does the frequency, structure and chemistry of dwarf galaxies require a more radical ingredient such as feedback from intermediate mass black holes?
\item
How are supermassive black holes formed in the early universe? Are intermediate mass black hole  seeds needed? Will gravity waves eventually probe their formation directly?
\item 
CMB: should we prioritize, if funding constraints require this,  B mode polarization  or  spectral distortions as the next new frontier of the very early universe? Or is the guarantee of primordial non-gaussianity a more fundamental and realistic goal?

\item Exploring the dark ages with 21cm at $z >20$ is the most promising but also the most challenging new frontier: is this feasible even at the most radio-quiet site known, on the far side of the moon?

\item Is inflation the correct description of the early universe?

\item Is the cosmological principle valid? This certainly motivated Einstein, Friedmann and Lemaitre. We should soon be able to test its validity, for example via all-sky surveys with the SKA.

\item Our cosmological model, and in particular our assessment of its contents such as dark matter and dark energy, assumes general relativity.  Is  general relativity  valid?
\end{itemize}

\subsection{Observations and experiment}
Finally I turn to a complementary wish list of observational questions. Cosmology is a field where observations of the natural phenomena, so sparse for millennia,  are now far ahead of theory. Here what one might expect observations to provide, someday.
\begin{itemize}

\item 
Spatially resolved spectroscopic mapping of molecular gas and star forming regions in galaxies near the peak of cosmic star formation at $z\sim 2.$
One could thereby attack the weakest link in galaxy formation theory, and understand the differences between early universe star formation,
when galaxies are youthful and gas-rich, with the present epoch that is witnessing  the inevitable fading of the bright lights of the universe.

\item 
Overcoming cosmic variance by targeting the scattering of cosmic microwave background photons  by hot gas in millions of galaxy clusters as probes of the local CMB quadrupole at redshifts up to redshifts $z\sim 2.$
One could then meaningfully address the large angular-scale anomalies in the cosmic microwave background, which are potential witnesses of a pre-inflationary past.

\item 
With sufficiently high precision  measurement of  CMB temperature fluctuations, determination of the spatial curvature of the universe on horizon scales becomes feasible.  One could then address one of the most fundamental questions in cosmology: how big is the universe?

\item 
Direct measurement of the expansion and acceleration of the universe by spectroscopy of distant objects at a resolving power of $10^6$ or greater. The second generation EELT CODEX spectrograph will be a first step, but one could do better with even larger telescopes in the future. 

\item 
 Resolution of Jeans mass neutral hydrogen scales in the dark ages, before the first stars formed. These are the ultimate building blocks of galaxies. Detection in absorption against the CMB  requires a low frequency radio interferometer on the far side of the moon with milliKelvin sensitivity at 50 MHz.

\item 
Extraction of  a primordial  non-gaussianity signal from the seed fluctuations that generated large-scale structure. Even the simplest, single-field inflation models predict such a signal at a level about 100 times below current limits. Our best bet here again may be a low frequency radio interferometer on the far side of the moon.

\item 
Resolution of the central engine in quasars and active galactic nuclei by microarc-second microwave imaging.
This also may require development of a lunar observatory in combination with a lunar satellite. 

\item
As for identification of dark matter, I see one promising strategy for exploring the final frontier of massive particle candidates. Let us assume  the dark matter consists of a weakly interacting particle.
The problem currently with all  astrophysics searches via indirect detection signatures is the lamp-post strategy: use of model predictions about electromagnetic signals for looking in the dark. Any Bayesian would admit that given any reasonable model priors, the probability of finding a signal would be low. History tells us that extending the energy frontier is a more fruitful approach towards discovering new phenomena. No guarantees of course, but I opt for a future 100 TeV proton collider as the next step towards exploring a promising regime where heavy dark matter particles may be hiding. One may not even need to seek much higher energies as  general unitarity arguments limit the mass  range to about 50 TeV  that one needs to search for hints of any new dark-matter related physics.

\end{itemize}
\vfill
\eject


\begin{thebibliography}{9}
\bibitem{tayler} Tayler, R \& Hoyle, F., Nature, 203, 1108 (1964)
\bibitem{burbidge} Burbidge, E. et al., RvMP, 29, 547 (1957)
\bibitem{coc} Coc, A., JPhCS, 665, 2001 (2016)
\bibitem{burles} Burles, S.,  NuPhA, 663, 861 (2000)
\bibitem{penzias} Penzias, A. \& Wilson, R., ApJ, 142, 419 (1965)
\bibitem{mather} Mather, J. et al., ApJ, 354, L37 (1990)
\bibitem{starobinsky}Starobinsky, A., Soviet Physics JETPL, 30, 682 (1979)
\bibitem{guth} Guth, A., PRD, 23, 347 (1981)
\bibitem{linde} Linde, A., PLB, 108, 393 (1982)
\bibitem{albrecht} Albrecht, A. \& Steinhard, P., PRL, 48, 1220 (1982)
\bibitem{sato} Sato, K., Phys. Lett., 33, 66 (1981)
\bibitem{mukhanov} Mukhanov, V.,  Soviet Physics JETP, 56, 258 (1982)
 \bibitem{rubin} Rubin, V. C., and W. K. Ford, Jr., ApJ. 159, 379 (1970)
 \bibitem{roberts} Roberts, M. S., and A. H. Rots, A\&A 26, 483 (1973)
 \bibitem{zwicky} Zwicky, F., Helvetica Physica Acta 6, 110 (1933)
 \bibitem{pagels} Pagels, H., and J. R. Primack, Phys.Rev.Lett. 48, 223 (1982)
 \bibitem{goodman} Goodman, M. \& Witten, E., PRD, 31, 3059 (1985) 
\bibitem{srednicki} Silk, J. \& Srednicki,  PRL, 53, 624 (1984)
\bibitem{sos} Srednicki, M., Olive, K. \& Silk, J., PRL, 55, 257 (1985)
 \bibitem{weinberg} Weinberg, S., Phys.Rev.Lett. 40, 223 (1978)
 \bibitem{wilczek} Wilczek, F., Phys.Rev.Lett. 40, 279 (1978)
 \bibitem{blumenthal} Blumenthal, G. et al., Nature, 313, 72 (1984)
 \bibitem{davis} Davis, M. et al., ApJ, 295, 371 (1985)
 \bibitem{bond83} Bond, J. R. and Szalay, A., ApJ, 274, 443 (1984)
 \bibitem{perlmutter} Perlmutter, S. et al., ApJ, 517, 565 (1999)
 \bibitem{riessa} Riess, A. et al., AJ, 116, 1009 (1998) 
 \bibitem{lange} Lange, A. et al., PRL, 63, 42001 (2001)
 \bibitem{silk} Silk, J., Nature, 215, 1155 (1967)
 \bibitem{silka} Silk, J., ApJ, 151, 459 (1968)
 \bibitem{wilson} Wilson, M. \& Silk, J., ApJ,  243, 14  (1981)
 \bibitem{peebles} Peebles, P. \& Yu, J., ApJ, 162, 815 (1970)
 \bibitem{eisenstein} Eisenstein, D. et al., ApJ, 633, 560 (2005)
 \bibitem{vittorio}  Vittorio, N. \& Silk, J., ApJ,  285, L39 (1984)
 \bibitem{bond}  Bond, J. R. \& Efstathiou, G., ApJ,  285, L45 (1984)
 \bibitem{keisler} Keisler, R. et al., ApJ, 743, 28 (2011)
 \bibitem{dunkley} Dunkley, J.  et al., ApJ, 379, 52 (2011) 
 \bibitem{sz} Sunyaev, R. \& Zeldovich, Ap\&SS, 7, 3 (1970)
  \bibitem {white} White, S. \& Rees, M., MNRAS,  183, 341 (1978)
 \bibitem{dekel} Dekel, A. \& Silk, J., ApJ, 303, 39 (1986)
  \bibitem{reess} Silk, J. \& Rees, M.,  A\&A, 331, L1 (1998)
\bibitem{planckXIII} Ade, P.  et al.,  A\&A 594, A13 (2016)
\bibitem{PlanckXLVI} Aghanim, N.  et al., A\&A in press, arXiv 1605.02985



\bibitem{hildebrandt2017} Hildebrandt, H., Viola, M., Heymans, C., et al., MNRAS, 465, 1454 (2017)

\bibitem{riess} Riess, A. et al., ApJ  in press, arXiv 1604.01424  (2016)

\bibitem{kamionkowskiP} Kamionkowski, M., Kosowsky, A. \& Stebbins, A., PRL, 78, 2058 (1977) 
\bibitem{seljak} Seljak, U. \& Zaldarriaga, M., PRL, 78, 2054 (1977)
\bibitem{hu94} Hu, W., Scott, D., \& Silk, J., ApJ, 430, L5  (1994)

\bibitem{pajer2012} Pajer, E., \& Zaldarriaga, M., Physical Review Letters, 109, 021302 920120  (2012)

\bibitem{sunyaev} Chluba, J. \& Sunyaev, R., MNRAS, 419, 1294 (2012)

\bibitem{percival}Percival, W., 
in Post-Planck Cosmology, Ecole de Physique des Houches, Les Houches, July 8-Aug 2, 2013, eds. B. Wandelt, C. Deffayet, P. Peter, Oxford: OUP, arXiv 1312.5490 (2016)
\bibitem{beutler16} Beutler, F. et al., MNRAS, 455, 1476 (2016)
\bibitem{euclid} Alemendola, L. et al.,  
Living Reviews in Relativity, 16, 270, updated in  arXiv:1606.00180 (2016)
\bibitem {candelas16} Candelas, P., Constantin, A. \& Mishra, C. , in press, arXiv 160206303 (2016) 
\bibitem{garcia12} Zumalacarregui, M.,  Garcia-Bellido, J. \& Ruiz-Lapente, R., JCAP 10, 009 (2012)
\bibitem{buchert16} Buchert T. et al., Class. Quantum Grav. 32,  215021 (2015)
\bibitem{mangano} Mangano, G.  et al., Nucl. Phys. B, 729, 221 (2005)
\bibitem{palanque} Palanque-Delabrouille, N. et al., JCAP, 11, 11 (2015)
\bibitem{gaskins16} Gaskins, J.,  in press arXiv160400014, Contemporary Physics (2016)
\bibitem{marrodan16} Marrodan Undagoitia,T. \& Rauch, L., J., Phys. G, 43, 013001 (2016)

\bibitem{cirelli2011} Cirelli, M., Corcella, G., Hektor, A., et al.,  JCAP, 3, 051 (2011)
\bibitem{daylan} Daylan, T. et al., Physics of the Dark Universe, 12, 1, arXiv 1402.6703 (2016)
\bibitem{lee} Lee, S. et al., PRL, 116, 1103 (2016)

\bibitem{dimauro} Di Mauro, M. et al., JCAP, 5, 31 (2016)


\bibitem{cirelli2015} Giesen, G., Boudaud, M., G{\'e}nolini, Y., et al., JCAP, 9, 023 (2015) 


\bibitem{kamionkowski} Griest, K.  \& Kamionkowski, M., PRL, 64, 615 (1990)
\bibitem{arkani16} Arkani-Hamed, N., Han, T.,  Mangano, M. \&  Wang, L., PhR, 652, 1 (2016)
\bibitem{shah} Shah, C. et al., in press ArXiv 160804, Ap J (2016)
\bibitem{profumo}  Jeltema, T. \& Profumo, S., MNRAS, 458, 392 (2016)
\bibitem{davis85} Davis, M. et al., ApJ, 292, 371 (1985)



\bibitem{palla83} Palla, F., Salpeter, E.~E., \& Stahler, S.~W., ApJ, 271, 632 (1983)



\bibitem{rees} Ostriker, J. \& Rees, M., MNRAS, 179, 541 (1977) 
\bibitem{silkb} Silk, J., ApJ, 211, 638 (1977)
\bibitem{rosdahl} Rosdahl, J. et al., MNRAS, in press, arXiv 1609.01296 (2015)
\bibitem{semenov} Semenov, V., Kravtsov, A. \& Genedin, N., ApJ, 826, 200 92016)
\bibitem{keller} Keller, B., Wadsley, J. \& Couchman, H., MNRAS, in press, arXiv 1604.08244 (2016)
\bibitem{salem} Salem, M., Bryan, G. \& Hummels, C. ApJ, 797, l8 (2014)



\bibitem{fattahi} Fattahi, A., MNRAS submitted, arXiv 1607.06479 (2016)
\bibitem{bland15} Bland-Hawthorn, J., Sutherland, R., \& Webster, D., ApJ, 807, 154 (2015)



\bibitem{merritt2016} Merritt, A., van Dokkum, P., Danieli, S., et al., arXiv:1610.01609  (2016)


\bibitem{vandokkum2015} van Dokkum, P.~G., Abraham, R., Merritt, A., et al., ApJL, 798, L45  (2015)



\bibitem{ji16} Ji A.~P., Frebel A., Chiti A., Simon J.~D., Nature, 531, 610  (2016)


\bibitem{fensch} Fensch, J. et al., MNRAS submitted, arXiv 1610.03877 (2016)

\bibitem{bose16} Bose, S., Hellwing, W.~A., Frenk, C.~S., et al., MNRAS, 455, 318 (2016)


\bibitem{marsh14} Marsh, D.~J.~E., \& Silk, J., MNRAS,  437, 2652 (2014)

\bibitem {hui16} Hui, L., Ostriker, J.~P., Tremaine, S., \& Witten, E., arXiv:1610.08297 (2016)


\bibitem{schneider16} Schneider, A., Trujillo-Gomez, S., Papastergis, E., Reed, D.~S., \& Lake, G., arXiv:1611.09362 (2016)

\bibitem{maccio13} Macci{\`o}, A.~V., Paduroiu, S., Anderhalden, D., Schneider, A., \& Moore, B.\  MNRAS, 428, 3715 (2013)

\bibitem{maccio12} Macci{\`o}, A.~V., Paduroiu, S., Anderhalden, D., Schneider, A., \& Moore, B.\ MNRAS, 424, 1105 (2012)

\bibitem{bozek15} Bozek, B., Marsh, D.~J.~E., Silk, J., \& Wyse, R.~F.~G.,\  MNRAS,450, 209 (2015)


\bibitem{abazajian2016} Abazajian, K.~N., Adshead, P., Ahmed, Z., et al., arXiv:1610.02743 (2016)


\bibitem{loeb05} Loeb, A. \& Zaldarriaga, M., PRD, 92, 3520 (2005)
\bibitem{maldacena} Maldacena, J., JHEP, 5, 13 (2003)
 \bibitem{renaux} Renaux-Petel, S., CRPhy, 16, 969 (2015)
 \bibitem{planckNG} Aghanim, N. et al., A\&A, 462, 4300 (2016)
\bibitem{munoz} Munoz, J., Ali-Hamoud, A. \& Kamionkowski, M., PRD, 91, 3521 (2015)
\end{thebibliography}
\end{document}